# MEASUREMENT OF THE TEMPORAL RESPONSE OF FERROELETRIC PHOTOCATHODES

M. Castellano, M. Ferrario, F. Tazzioli, INFN-LNF; L. Catani, INFN-ROMA2;
L. Giannessi, ENEA-CRE, I. Boscolo, S. Cialdi, M. Valentini, University and INFN Milano.

*Abstract*

Various photoemissive materials, among which ferroelectric ceramics are tested as photo emitters at INFN Frascati Laboratories. In order to understand the physical processes involved in the emission and to characterize them as photocathodes for use in linac injectors, it is important to measure the temporal shape of the emitted current. As the duration of the laser pulse is 25 ps, the required resolution is of the order of few ps. An apparatus has been set up for the purpose, consisting of a 30 kV electron gun, a microwave transverse deflecting cavity which translates the temporal distribution of the electron bunch into a spatial one, a fluorescent screen on which the deflected beam traces a sector of a circle and various focusing and charge measuring items. The image on the screen is detected via a CCD camera and a frame grabber. We describe the performance of the apparatus and some preliminary temporal distribution measurements.

## 1 INTRODUCTION

The measurement of the time correlation between laser pulse and emitted current from a photocathode is important to characterize the cathode itself and the suitability of the generated electron pulse for further acceleration in injectors. In our case the photon pulse is obtained from a mode locked Nd-Yag laser and has a duration of 25 ps. The cathodes to be tested are made of either ferroelectric PLZT ceramics [1] or doped Diamond [2]. The ceramics are expected to exhibit a multiphoton response starting from the green wavelength range (532 nm) while the diamond samples start emitting in the UV range. The measuring system has been initially set up for the green light (second harmonic of the laser) but can be readily adapted to UV radiation. The ceramic cathodes are disks of 16 mm diameter, 0.7 to 1 mm thickness, provided with a uniform metallic film electrode on the rear side and with a grating interconnected by an external ring on the front (emitting) side. This configuration allows the application of a bias to improve the emission efficiency [3]. For the same reason the sample can also be heated by a resistive wire. The cathodes are rugged and need no special precautions in handling. The vacuum requirements are dictated by the high voltage used in the gun.

## 2 EXPERIMENTAL SYSTEM

The bunch length is measured by a known technique [4] similar to that used in streak cameras. The beam, generated by photo excitation and accelerated to about 30 keV in a DC gun, is deflected by two sinusoidal crossed magnetic fields, in time quadrature with each other, which impress a transverse momentum to the various slices of the bunch. The image seen on a fluorescent screen downstream of the deflector is a sector of a circle whose length, divided by the full circumference, gives directly the bunch length in units of the RF period. A sketch of the system is shown in Fig.1.

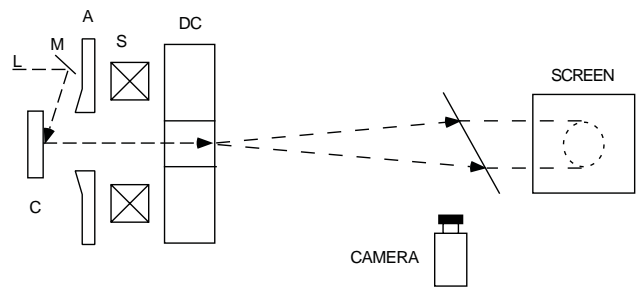

Figure 1: Measuring system. C=cathode, L=laser beam, M= mirror, A=anode, S=solenoid, DC= deflecting cavity.

The deflecting fields are created by exciting the TM110 mode in a microwave resonator powered at 2450 MHz by two loop antennas spaced by 90 degrees on the azimuth. The loaded Q of the Aluminum cavity in matched condition is 5000 and the microwave power required to impart a deflection of 3 cm at a distance of 40 cm to a beam having an energy of 50 keV is 250 W for each antenna. The duration of the microwave pulse is determined by the jitter in the starting time of the laser pulse with respect to the trigger, which is about 4 μs. An RF duration of 10 μs has been chosen. A sketch of the microwave system is shown in Fig. 2.

The microwave signal does not need to be synchronized with the laser pulse because varying phases of the bunch with respect to the RF waveform produce only an azimuth shift of the image, which can be instantaneously grabbed pulse by pulse. This means that the microwave generator can be an oscillator (magnetron), which is much cheaper than a klystron amplifier.

The diagnostic devices consist of the microwave cavity itself, a wall current monitor and a fluorescent ceramic screen plus a CCD camera with frame grabber. The signals induced in the monopole modes of the cavity can be used as a very sensitive indicator of the presence of the beam in the initial setting up of the instrument. The wall current monitor can be used to estimate the bunch charge and the bunch length of very long bunches (exceeding the RF period).

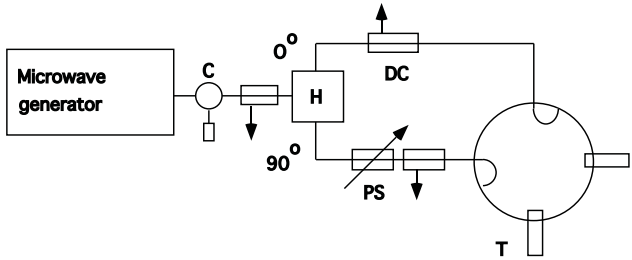

Fig 2. Sketch of the microwave system. C= circulator, H= hybrid junction, PS= phase shifter, T= tuner, DC= directional coupler.

## 3 OPERATION OF THE SYSTEM

The system has been set operating with green wavelength (532 nm). This makes laser beam alignment easier but brings serious problems of image detection because the emission efficiency of the ceramic cathodes at this wavelength is still very low. The laser signal to obtain tens of pC of bunch charge is of the order of milliJoules. The green light diffused by the cathodes illuminates the target and overwhelms the fluorescence induced by the beam. We have solved this problem by introducing a low pass filter in front of the camera, taking advantage of the fact that the Chromium doped ceramic employed emits substantially red light. In Fig. 3 is shown a picture of the beam taken with the frame grabber.

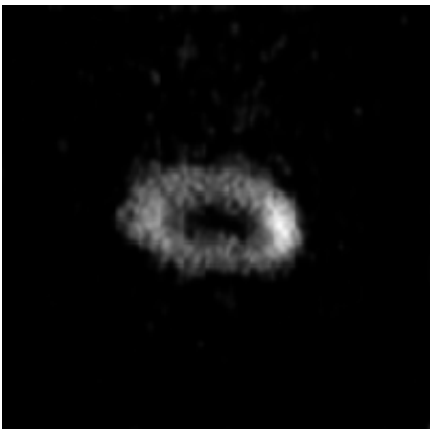

Figure 3: Image of the beam on the screen.

The shape of the image, for a continuous beam, should be an ellipse due to the screen being bent at 45 degrees with respect to the electron beam axis, but can also be deformed by an imperfect balance of the two microwave channels. The resolution of the time measurement is given by the thickness of the trace which is mainly the dimension of the beam spot focused by the solenoid. The minimum dimension of the spot is determined by the beam emittance and space charge.

In Fig. 4 is shown the image intensity distribution along the perimeter of the elliptical figure.

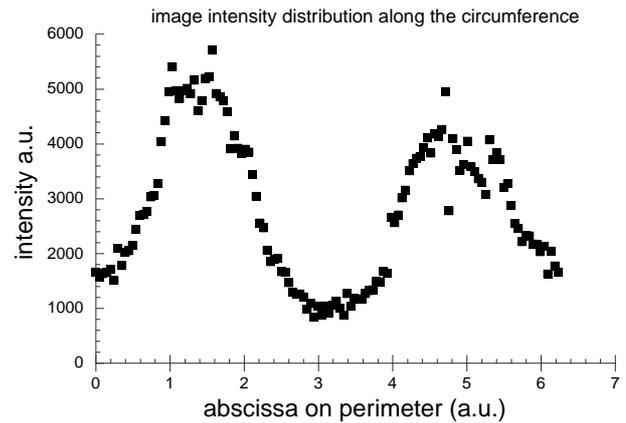

Figure 4: Image intensity distribution.

One notices that the beam trace fills the whole perimeter of the ellipse. This means that in this case the bunch length is very large. In fact due to the high power density of the laser beam (3 mJ on a laser spot of about 2 mm diameter) the cathode surface is very close to breakdown. The image of the head of the bunch is superimposed to a periodic pedestal due to the tail of the bunch and to the imperfect balance of the sinusoidal deflecting fields. In order to see the perimeter partially filled (and therefore only one bump on the figure) one should decrease the laser power and beam intensity, but one cannot go below the sensitivity threshold of the ceramic screen. We plan to install a higher efficiency cathode and to employ an intensified camera. A long bunch can anyhow be useful to balance the deflecting fields by minimizing the pedestal bumps.

It is to be noticed that with this system we measure the bunch length at the deflecting cavity location, which is at 20 cm from the cathode. This length, for a given laser pulse length, depends on gun voltage and space charge. In Figures 5 and 6 are shown the evolutions of bunch length from the cathode to the deflection cavity for different bunch charges, as obtained with the HOMDYN code [5].

Space charge effects result in a bunch longer than the initial laser pulse length $<L>=8$ ps, even in the low charge limit of 1 pC. For the same reason we think that a direct measurement of the electron distribution exiting from the surface of the cathode is difficult by this technique, unless low charge and high voltage are applied, nevertheless one can obtain some useful information about the cathode temporal response by comparing measurements with computations.

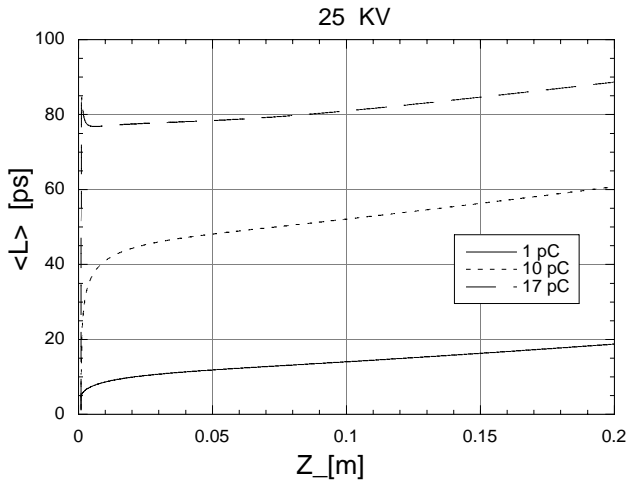

Figure 5: Evolution of bunch length from cathode to deflector by 25 kV voltage applied. Maximum extracted charge in this case is 17 pC

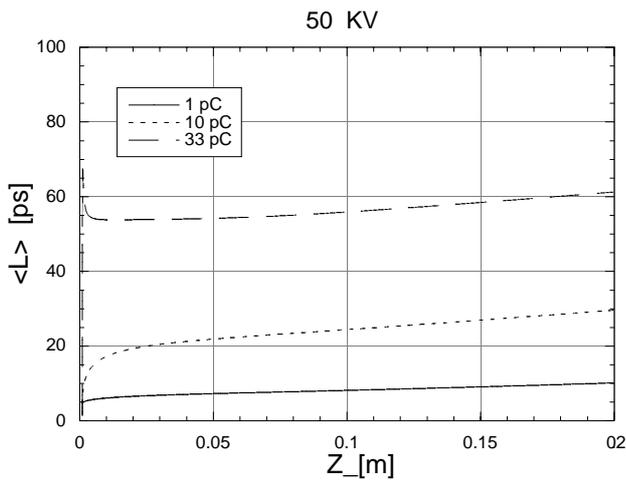

Figure 6: Evolution of bunch length from cathode to deflector by 50 kV voltage applied. Maximum extracted charge in this case is 33 pC.

## 4 CONCLUSIONS

An instrument which is able to measure the bunch lengths of electrons emitted by a photocathode with a resolution of a few picoseconds has been successfully set operating. Only very preliminary results have been obtained but we plan to proceed in the near future to calibrate accurately the instrument and to perform bunch length measurements on the various rugged photocathodes that are being studied in our laboratory.

## AKNOWLEDGMENTS

We wish to thank the technicians L. Cacciotti and R. Sorchetti whose work has been essential to obtain these results.